\documentclass[%
reprint,
superscriptaddress,
nofootinbib,
nobibnotes,
amsmath,amssymb,
aps,
prl,
floatfix,
]{revtex4-2}

\usepackage{epsfig}
\usepackage{bm}
\usepackage{amssymb}
\usepackage{amsmath}
\usepackage{color}
\usepackage{xcolor}
\usepackage{hyperref}
\hypersetup{colorlinks, linkcolor = [rgb]{0, 0, 0.5}, citecolor = [rgb]{0,0.0,0.5}, urlcolor = [rgb]{0,0.0,0.5}}
\usepackage{multirow}
\usepackage{slashed}
\usepackage[multiple]{footmisc}
\usepackage{dcolumn}
\usepackage{ulem}
\usepackage[left, pagewise]{lineno} %

\allowdisplaybreaks[4]

\mathchardef\mhyphen="2D

\begin{document}
	
\title{Suppression of Spin Transfer to \texorpdfstring{$\Lambda$}{} Hyperon in Deep-Inelastic Scattering}
	
\newcommand*{\SDU}{Key Laboratory of Particle Physics and Particle Irradiation (MOE), Institute of Frontier and Interdisciplinary Science, Shandong University, Qingdao, Shandong 266237, China}\affiliation{\SDU}
\newcommand*{\SCNT}{Southern Center for Nuclear-Science Theory (SCNT), Institute of Modern Physics, Chinese Academy of Sciences, Huizhou 516000, China}\affiliation{\SCNT}
	
\author{Xiaoyan Zhao}
\affiliation{\SDU}
	
\author{Zuo-tang Liang}
\affiliation{\SDU}
	
\author{Tianbo Liu}
\email[Contact author: ]{liutb@sdu.edu.cn}
\affiliation{\SDU}\affiliation{\SCNT}
	
\author{Ya-jin Zhou}
\email[Contact author: ]{zhouyj@sdu.edu.cn}
\affiliation{\SDU}

\begin{abstract}
		
We investigate $\Lambda$ production in semi-inclusive deep-inelastic scattering using a polarized lepton beam and find that the spin transfer is significantly suppressed by target fragmentation. As further demonstrated by a model estimation, experimental data can be well described once the target fragmentation is taken into account, which alleviates the tension with calculations solely based on current fragmentation. Our findings suggest that, at the energies of existing fixed-target experiments, the separation of current and target fragmentation regions is not distinct. Spin transfer as well as other spin effects offers a sensitive probe into the origin of the produced hadron.
        
\end{abstract}

\maketitle
	
	
{\it Introduction}---The $\Lambda$ polarization in high-energy scattering experiments has received great interest since the first observation of spontaneous transverse polarization in 1976~\cite{Bunce:1976yb}, which was not expected in perturbative QCD(pQCD)~\cite{Kane:1978nd}. This landmark event sparked extensive experimental measurements and theoretical explorations, providing invaluable insights into understanding the spin effects in hadronization processes, nucleon spin structures, nuclear matter properties under extreme conditions in heavy-ion collisions, and hadron spectroscopy. 
	
Owing to the self-analyzing weak decay~\cite{Lee:1957qs}, $\Lambda$ polarization is a powerful tool for investigating spin-dependent fragmentation functions (FFs)~\cite{Metz:2016swz,Chen:2023kqw}, and numerous measurements have been carried out in electron-positron annihilations~\cite{ALEPH:1996oew,OPAL:1997oem,Belle:2018ttu}, lepton-hadron deep-inelastic scatterings ~\cite{E665:1999fso,HERMES:1999buc,NOMAD:2000wdf,ZEUS:2006ugq,HERMES:2006lro,COMPASS:2009nhs,COMPASS:2021bws}, and hadron-hadron scatterings~\cite{STAR:2009hex,ATLAS:2014ona,STAR:2018fqv,STAR:2018pps,STAR:2023hwu}. Since $\Lambda$ is an isoscalar hyperon with its spin dominated by the valence strange quark~\cite{Gell-Mann:1964ewy,Zweig:1964ruk}, the polarization of $\Lambda$ produced in semi-inclusive deep-inelastic scattering (SIDIS) is expected to have a strong sensitivity to the strange sea of the nucleon~\cite{Ellis:1995fc, Lu:1995np, Brodsky:1996hc, Jaffe:1996wp, Ma:2000uv, Ellis:2007ig, Chi:2014xba, Du:2017nzy}. Unlike the electron-positron annihilation process, in which all hadrons are generated from the produced quark-antiquark pair, the identified final-state hadron in SIDIS may either come from the struck quark, referred to as the current fragmentation (CF)~\cite{Berman:1971xz,Collins:1981uw,Metz:2016swz}, or come from the remnant of the initial-state nucleon, referred to as the target fragmentation (TF)~\cite{Trentadue:1993ka,Berera:1995fj,Grazzini:1997rw,deFlorian:1997wi, Grazzini:1999vz}. While most studies focused on the CF mechanism~\cite{Meng:1991da,Meng:1995yn,Nadolsky:1999kb,Ji:2004xq,Ji:2004wu,Koike:2006fn,Collins:2011zzd,Collins:2017oxh}, the theoretical framework for the TF has also developed and refined in recent years~\cite{Ceccopieri:2007th,Anselmino:2011ss, Chai:2019ykk, Yang:2020sos, Chen:2021vby, Chen:2023wsi, Chen:2024brp}.
	
To understand the $\Lambda$ polarization in SIDIS, it is important to trace the production mechanism, leading to a challenging task in experimental analysis to separate CF and TF regions. In an ideal limit, when the virtual photon carries infinite momentum, one can imagine that the hadrons from CF are moving nearly around the photon direction and those from TF are likely at the end of the direction of the nucleon momentum. This feature was observed at HERA~\cite{ZEUS:1993vio}. However, such a rapidity gap is not observed at existing fixed-target experiments or early neutrino scattering experiments~\cite{Ammosov:1980zs, Birmingham-Bonn-CERN-London-Munich-Oxford:1985ret, WA59:1991cop, Jones:1992bm, NOMAD:2001iup, E632:1994njp, EuropeanMuon:1984ucw}, and events from the CF region and TF region all mix together.
Although some practical criteria were proposed~\cite{Berger:1987zu,Mulders:2000jt,Boglione:2016bph,Boglione:2019nwk}, there is, in fact, no clear border between the CF and the TF regions.
	
Instead of creating another criterion, we propose in this Letter to take advantage of the longitudinal spin transfer $D_{LL}$ as a sensitive observable to the origin of the produced $\Lambda$. By examining the spin-flavor structure, we find the $D_{LL}$ to $\Lambda$, which has been measured by HERMES~\cite{HERMES:2006lro,Belostotski:2011zza}, COMPASS~\cite{COMPASS:2009nhs}, and CLAS~\cite{McEneaney:2022bsf,McEneaney:2023s}, will be significantly suppressed if a sizable fraction of events are from the TF.
To quantitatively demonstrate this mechanism, we perform a model estimation.
Our results indicate that the spin transfer is highly suppressed for measurements at low energies even if only the leading TF channel is taken into account, and this effect reduces at higher energies.
Therefore, the spin suppression effect from TF not only alleviates the tension between data and theoretical predictions with CF, but also provides a new perspective to explore the hadronization mechanism.
This mechanism can be further tested at current experiments at Jefferson Lab (JLab)~\cite{Dudek:2012vr} and future electron-ion colliders~\cite{Accardi:2012qut,AbdulKhalek:2021gbh,Anderle:2021wcy}.

	
{\it Spin transfer}---We consider $\Lambda$ production in the SIDIS process using a polarized lepton beam and an unpolarized nucleon target:
\begin{align}
    l(\ell) + N(P) \to l(\ell') + \Lambda(P_\Lambda ) + X,
\end{align}
where the labels in parentheses represent the momenta of corresponding particles, and $X$ stands for the undetected hadronic system. With one-photon-exchange approximation, the differential cross section is given by~\cite{Boer:1999uu, Bacchetta:2006tn, Yang:2016qsf, Zhao:2024zpy},
\begin{align}
    &\frac{d\sigma}{dxdydz}= \frac{4\pi^2\alpha^{2}}{xyQ^{2}}\frac{y^2}{2(1-\varepsilon)}\left(1+\frac{\gamma^2}{2x} \right)
	\notag \\
	&\ \ \times\left[
	F_{UU}^U(x,Q^2,z) + \lambda_e \lambda_\Lambda \sqrt{1 - \varepsilon^2} F_{LU}^L(x,Q^2,z)
	\right],
\end{align}
where $\alpha$ is the electromagnetic fine structure constant, $\lambda_e$ represents the helicity of the lepton beam, and $\lambda_\Lambda$ represents the polarization of produced $\Lambda$ hyperons. Kinematic variables are defined as
\begin{align}
	Q^2 &= -q^2 = -(\ell - \ell')^2,
	\quad
	x = \frac{Q^2}{2P\cdot q},
	\notag\\
	y &= \frac{P\cdot q}{P\cdot \ell},
	\quad
	z = \frac{P\cdot P_\Lambda}{P\cdot q},
	\quad
	\gamma = \frac{2 x M}{Q},
	\notag\\
	\varepsilon &= \frac{1-y-\frac{1}{4} \gamma^2 y^2}{1-y+\frac{1}{2} y^2+\frac{1}{4} \gamma^2 y^2}.
\end{align}
The structure functions $F_{UU}^U$ and $F_{LU}^L$ correspond to unpolarized and longitudinally polarized $\Lambda$ production, respectively, as indicated by the superscripts. The subscripts indicate the polarizations of the lepton and the nucleon.
By flipping the spin of the lepton beam, one can measure the ratio of the two structure functions:
\begin{align}
    D_{LL}^\Lambda = \frac{F_{LU}^L(x,Q^2,z)}{F_{UU}^U(x,Q^2,z)},
\label{eq:DLL}
\end{align}
referred to as the longitudinal spin transfer.
	
In this process, a polarized virtual photon is emitted from the lepton beam. The struck quark in the nucleon is then polarized after absorbing the photon. If assuming the CF, in which the detected $\Lambda$ is generated from the fragmentation of the struck quark, the spin orientation of the $\Lambda$ may be related to the polarization of the struck quark. 
This correlation is characterized by the polarized FF $G_{1q}^\Lambda(z)$, which represents the density of longitudinally polarized $\Lambda$ carrying a fraction $z$ of momentum of the parent polarized quark with flavor $q$.
According to this mechanism, one can factorize the $F_{LU}^L$ into the convolution of $f_{1q}(x)$ and $G_{1q}^\Lambda(z)$, where $f_{1q}(x)$ is the unpolarized parton distribution function (PDF) of the nucleon, describing the parton density of flavor $q$ that carries a fraction $x$ of the nucleon momentum. Similarly, the $F_{UU}^U$ is factorized into the convolution of $f_{1q}(x)$ and the unpolarized FF $D_{1q}^\Lambda(z)$. Leading-order expressions are given by
\begin{align}
	F_{UU}^U &= \sum_q e_q^2 x f_{1q}(x) D_{1q}^\Lambda(z),
	\label{eq:FUU_CF}\\
	F_{LU}^L &= \sum_q e_q^2 x f_{1q}(x) G_{1q}^\Lambda(z),
	\label{eq:FUL_CF}
\end{align}
where $e_q$ is the fractional electric charge of the quark.
	
For the TF, in which the detected $\Lambda$ is from the remnant of the nucleon, the hadronization process is formally described by the fracture function (FrF) $M^U_{q}(x,\zeta)$.
Here $\zeta$ is the momentum fraction of the nucleon carried by the final-state $\Lambda$, and $x$ is the momentum fraction carried by the struck quark $q$, the same meaning as in the PDF.
Although the detected $\Lambda$ does not directly learn the struck quark information, its polarization is, in general, still related to the spin of the struck quark, because the spin states of the struck quark and the remnant of the nucleon, referred to as the spectator, are correlated as constrained by the nucleon wave function.
In the reaction considered here, the polarized photon has a preference to pick quarks at certain spin states required by the helicity conservation, and, thus, the $\Lambda$ generated from the nucleon remnant can, in principle, have nonzero polarization, even if the virtual photon polarization is not directly transferred to the nucleon remnant.
Such correlation is formally characterized by the polarized FrF $\Delta M^L_{q}(x,\zeta)$. Following the convention of FrFs in~\cite{Anselmino:2011ss}, the structure functions within the TF are expressed as
\begin{align}
	F_{UU}^U &= \Big|\frac{\partial \zeta}{\partial z}\Big| \sum_q e_q^2 x  M_{q}^U (x,\zeta), \\
	F_{LU}^L &= \Big|\frac{\partial \zeta}{\partial z}\Big| \sum_q e_q^2 x  \Delta M^L_{q} (x,\zeta),
\end{align}
where $|\partial \zeta / \partial z|$ is the Jacobian.
	
To estimate the effect of TF on the longitudinal spin transfer to $\Lambda$, we start from the analysis of kinematic regions. For convenience, we take the reference frame in which the nucleon and the virtual photon are head on, with the nucleon moving backward. After the collision, the struck quark acquires large energy from the virtual photon and moves forward, while the nucleon remnant keeps moving backward. Then the hadrons detected in the forward region are likely from the CF and those in the backward region are likely from the TF. If further considering the color neutralization process, which requires radiation of partons to combine into hadrons, one will expect at the high-energy limit that most hadrons are generated in the CF region, because the energy is mostly transferred to the struck quark. 
	
However, at medium energies, such as the experiments at JLab, HERMES, and COMPASS, no clear isolation of CF and TF regions is observed, and one should, in principle, take both mechanisms into account. Because limited phase space is given to the struck quark in this case, a quark of flavor $u$, $d$, or $s$ will have a better chance to produce a $\Lambda$, which has the valence component $|uds\rangle$. It is often referred to as favored fragmentation channels. If also considering the parton densities of the nucleon, the $u$ and $d$ channels would be more dominant. Hence even for leading CF channels, the struck quark needs to generate at least two pairs of quarks to form a $\Lambda$, and, correspondingly, the FF has the falloff behavior in powers of $(1-z)$ at large $z$. According to~\eqref{eq:FUU_CF} and~\eqref{eq:FUL_CF}, $D_{LL}$ is characterized by the ratio between the PDF-weighted sum of $G_{1q}^\Lambda$ and that of $D_{1q}^\Lambda$ if only CF contributions are taken into account. While it generally relies on the knowledge of FFs, the pQCD helicity retention~\cite{Farrar:1975yb,Brodsky:1994kg} predicts at the $z\to1$ limit  $G_{1q}^\Lambda$ and $D_{1q}^\Lambda$ tend to be the same, and, thus, $D_{LL} \to 1$.
	
For TF, the $\Lambda$ is produced from the nucleon remnant, which contains a bunch of partons. To match the flavor of $\Lambda$, a $uds$ quark cluster undoubtedly has the best chance, but the probability to find such a cluster in the nucleon is extremely low. We note that the possible existence of polarized intrinsic strange sea~\cite{Ellis:1995fc,Brodsky:1996hc} may contribute to the $\Lambda$ polarization via this channel, which is beyond the scope of this Letter. Then, one may consider the quark pair of $ud$, $us$, or $ds$, which needs to generate only one additional quark to form the $\Lambda$. Since the $ud$ pair exists in the valence component of the nucleon, it is expected to be the dominant channel. As $\Lambda$ is an isoscalar hyperon, the $ud$ pair should form a scalar in its valence component.
Therefore, the leading TF channel will generate unpolarized $\Lambda$ and, consequently, reduce the $D_{LL}$ value.
Comparing the dominant CF and TF channels, one can find that the TF requires less phase space for radiation. So the suppression of the spin transfer $D_{LL}$ by TF will be significant for experiments at relatively low energies.

	
{\it Numerical estimation}---To quantitatively demonstrate the suppression of spin transfer by TF, we need the knowledge of FFs and FrFs. These are essentially nonperturbative quantities, and first-principle calculations are still unavailable. For FFs $D_{1q}^\Lambda$ and $G_{1q}^\Lambda$, we take the parametrization~\cite{deFlorian:1997zj}. Since the fit was obtained by analyzing $e^+e^-$ annihilation data, which contain no hadrons in the initial state, we do not need to worry about the contamination from TF for the extracted FFs.
	
As the pQCD global analysis of FrFs is still missing and our main focus is to demonstrate the importance of TF in understanding the spin transfer, instead of a precise description, we adopt a model calculation. Here, we choose the quark-diquark model, which has been extensively utilized in phenomenological studies and proven successful in quantitatively describing many physical quantities, such as form factors, unpolarized and polarized PDFs, transverse-momentum-dependent PDFs, generalized PDFs, FFs, and Wigner distributions. In this model, the nucleon is viewed as an active quark that is probed by the virtual photon and a diquark as the remnant of the nucleon. The diquark is either a scalar or an axial vector, which can be derived from the SU(6) spin-flavor wave function of the nucleon, and similarly for the $\Lambda$.
	
We follow the formalism in Ref.~\cite{Bacchetta:2008af} by introducing effective quark-diquark-baryon vertex:
\begin{align}
	\Upsilon_s = i g_s {\bf 1},
	\quad
	\Upsilon_a^\mu = i \frac{g_a}{\sqrt{2}}\gamma^\mu \gamma_5,
\end{align}
where $s$ and $a$ stand for the scalar diquark and the axial-vector diquark, respectively. The Gaussian-type form factors are chosen for the couplings as $g_{s(a)}\to g_{s(a)} \exp\{[(p^2 - m_q^2) - (p_{s(a)}^2 - M_{s(a)}^2)]/\lambda_{s(a)}^2\}$, where $p$ and $p_{s(a)}$ are the momenta of the quark and diquark, respectively, $m_q$ and $M_{s(a)}$ are quark and diquark masses, respectively, and $\lambda_{s(a)}$ is a cutoff parameter. The quark masses are set as $m_u = m_d = 0.3\,\rm GeV$ and $m_s = 0.5\,\rm GeV$, and other parameters $M_s = 1.2\,\rm GeV$, $M_a = 1.3\,\rm GeV$, $\lambda_s = 2.9\,\rm GeV$, $\lambda_a = 1.8\,\rm GeV$ are tuned to match valence quark PDFs $f_{1q}(x)$ in the JR14 parametrization~\cite{Jimenez-Delgado:2014twa}.
	
\begin{figure}[htp]
    \centering
    \includegraphics[width=0.9\linewidth]{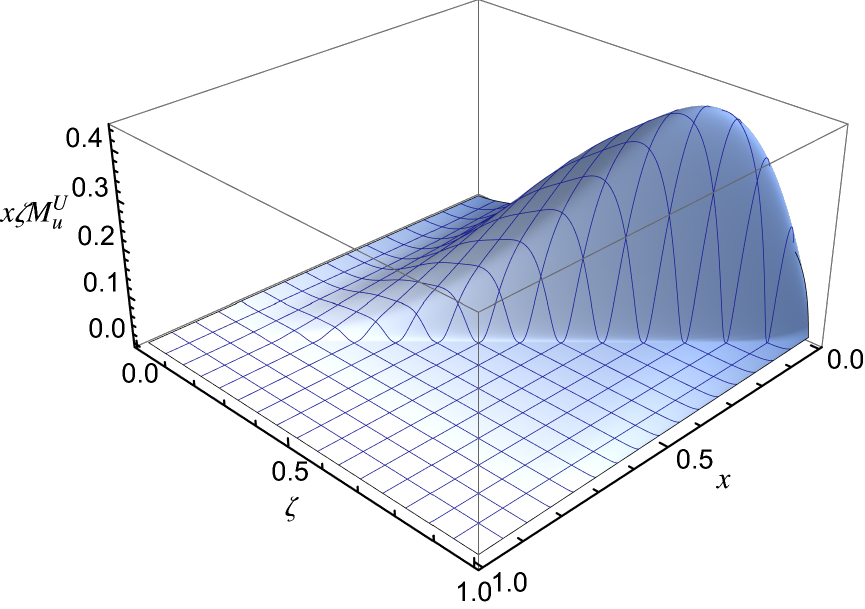}
	\caption{Model results of the FrF $M_{u}^U(x,\zeta)$ of the proton with a $u$ quark as the struck quark.}
	\label{fig:MUU}
\end{figure}

With the correlation matrix for the FrFs~\cite{Anselmino:2011ss}, we obtain the expression of the transverse-momentum-dependent FrF as
\begin{align}
	M^{U}_{u}&
	=\frac{3g_{sN}^2 g_{s\Lambda}^2 x^2 \left[ (m_u+x M)^2+\boldsymbol p_\perp^2\right] }{4 (2\pi)^6 \zeta^2(1-\zeta-x)^2 (p^2-m_u^2)^2} \\
	&\times \frac{\left[ (1-x-\zeta)M_\Lambda -\zeta m_{s} \right]^2+\left[ (1-x) {\bm P}_{\Lambda \perp}+\zeta {\bm p}_\perp \right]^2 }{ \left[ x(1-x)M^2-x M_s^2-(1-x)p^2-\boldsymbol{p}_\perp^2 \right]^2 }, \notag 
	\label{eq:MUU}
\end{align}
where ${\bm p}_\perp$ and ${\bm P}_{\Lambda\perp}$ are the transverse momenta of the active quark and the $\Lambda$ with respect to the nucleon, respectively, and $g_{sN}$ and $g_{s\Lambda}$ are form factors in the quark-diquark-nucleon and the quark-diquark-hyperon vertices, respectively, as introduced above. 
The transverse-momentum-integrated result of $M_{u}^{U}(x,\zeta)$ with a struck $u$ quark is shown in Fig.~\ref{fig:MUU}.
Here, we have considered the valence component of the proton. With the struck quark being a $u$ quark, the remnant is a $ud$ diquark, which contributes to the leading TF channel according to the previous analysis.

\begin{figure*}[htp]
	\centering
    \includegraphics[width=0.9\linewidth]{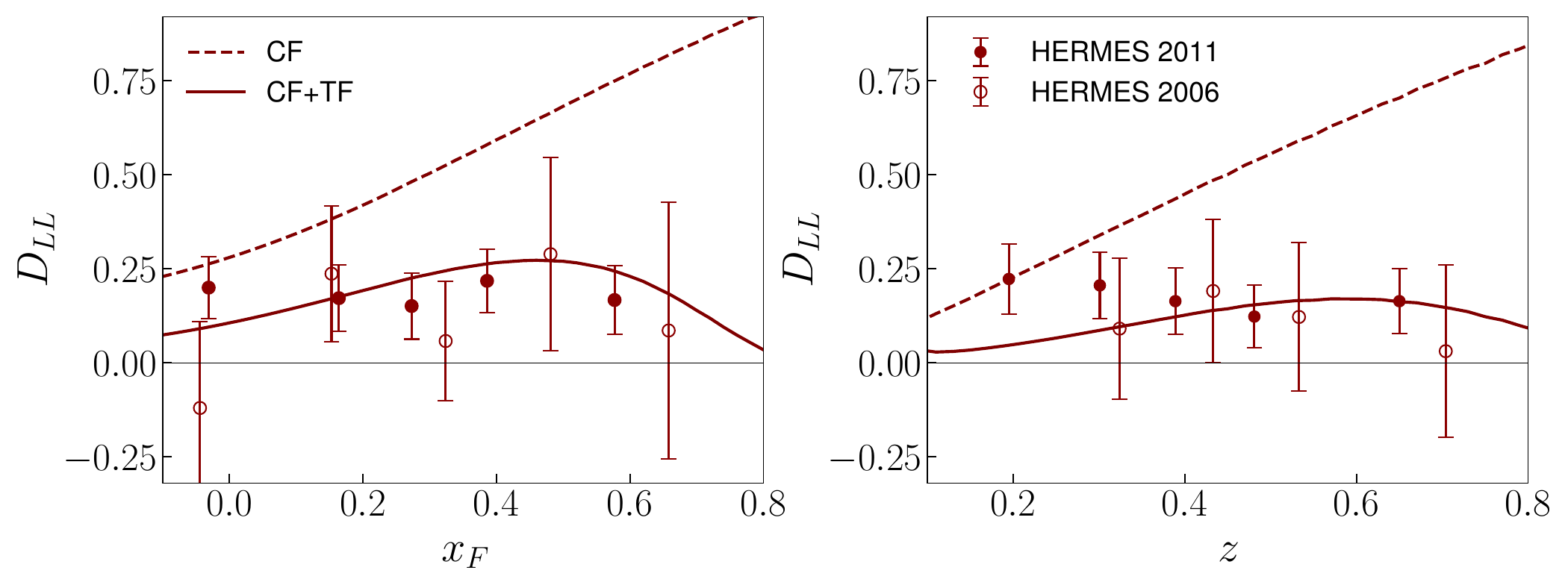}
	\caption{Longitudinal spin transfer $D_{LL}$ in comparison with HERMES data~\cite{HERMES:2006lro,Belostotski:2011zza}. The dashed curves include only CF contributions, and the solid curves include both CF and TF contributions. The curves are evaluated at the average values of  $x=0.088$ and $Q^2 = 2.4\,\rm GeV^2$.
    We note that the data here were analyzed with $\alpha_\Lambda = 0.642  \pm 0.013$ and are not rescaled by the latest value $\alpha_\Lambda = 0.746 \pm 0.009$~\cite{ParticleDataGroup:2024cfk}.}
	\label{fig:compare_hermes}
\end{figure*}

\begin{figure*}[htp]
	\centering	
    \includegraphics[width=0.9\linewidth]{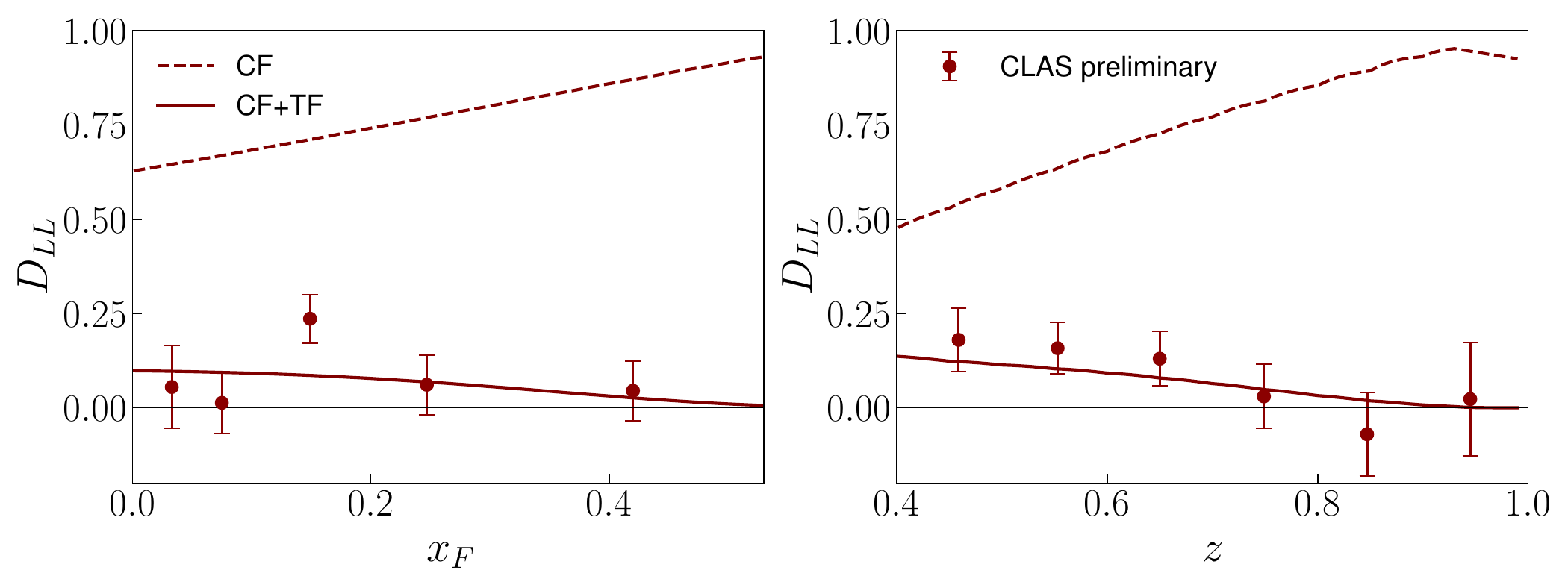}
	\caption{Longitudinal spin transfer $D_{LL}$ in comparison with CLAS preliminary data~\cite{McEneaney:2022bsf, McEneaney:2023s}. The dashed curves include only CF contributions, and the solid curves include both CF and TF contributions. The curves are evaluated at the average values of $x=0.25$ and $Q^2=2.13\,\rm GeV^2$.
    We note that the data here were analyzed with $\alpha_\Lambda = 0.732 \pm 0.014$ and are not rescaled by the latest value $\alpha_\Lambda = 0.746 \pm 0.009$~\cite{ParticleDataGroup:2024cfk}.}
	\label{fig:compare_claspre}
\end{figure*}

Then we calculate the $D_{LL}$, in~\eqref{eq:DLL}, in which the structure functions are integrated over the transverse momentum. For the transverse momentum dependence of PDFs and FFs, we use the common Gaussian parametrization in Refs.~\cite{Anselmino:2013lza,Callos:2020qtu}.
In Fig.~\ref{fig:compare_hermes}, the results are shown in comparison with HERMES data~\cite{HERMES:2006lro,Belostotski:2011zza}.
One can observe that the results with only CF contributions severely deviate from the data. The increasing behavior at large $x_F$ and $z$ is expected from the pQCD helicity retention prediction for $G_{1q}^\Lambda/D_{1q}^\Lambda$. However, this trend is not observed in the measurement, even at very large $x_F$ and $z$. When the TF is taken into account, the results can match the data well. The suppression of $D_{LL}$ by TF contributions is significant. One may notice that at large $x_F$ and $z$, where the detected $\Lambda$ is along the virtual photon direction, the suppression effect is even more significant. It seems in contradiction with the intuitive picture. The reason is that in the relatively low-energy experiment the struck quark has very limited phase space to shower, especially when the measured hadron carries a large fraction of momentum. Thus, the CF contribution is also inhibited at very large $x_F$ and $z$ as presented in the Supplemental Material~\cite{SupplementalMaterial}. Furthermore, at this limit, the spin transfer from CF tends to $1$, while that from TF stays around $0$. The huge difference between the values predicted from the two mechanisms makes the suppression effect more visible.
	
In Fig.~\ref{fig:compare_claspre}, the results are compared with JLab CLAS preliminary data~\cite{McEneaney:2022bsf, McEneaney:2023s}. As can be observed, the tension between CLAS data and theoretical predictions with only CF contributions is more serious. This is consistent with our analysis, because the energy of the CLAS experiment is even lower than that of HERMES. Therefore, one can expect more significant effect from the TF. Once the TF is taken into account, the data are well described.
	
\begin{figure}[htp]
	\centering
    \includegraphics[width=0.9\linewidth]{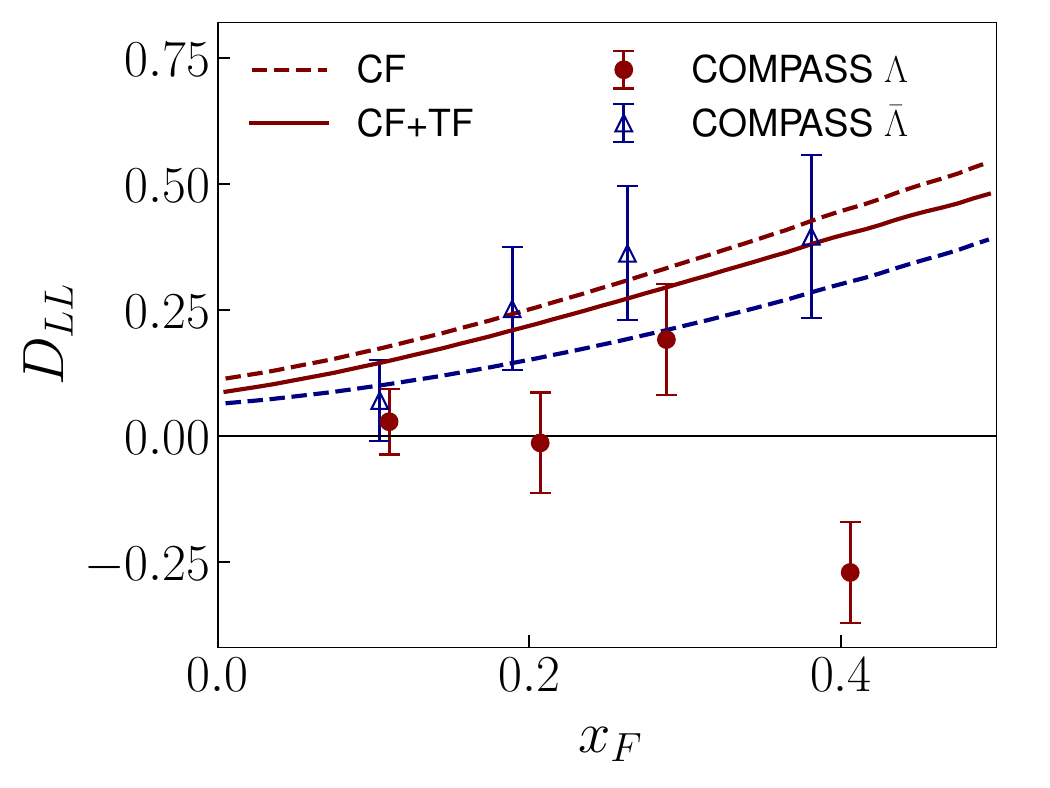}
	\caption{Longitudinal spin transfer $D_{LL}$ in comparison with COMPASS data~\cite{COMPASS:2009nhs}. The dashed curves include only CF contributions, and the solid curves include both CF and TF contributions. Theoretical curves are evaluated at $x=0.03$ and $Q^2=3.7\,\rm GeV^2$.
    We note that the data here were analyzed with $\alpha_\Lambda = 0.642  \pm 0.013$ and are not rescaled by the latest value $\alpha_\Lambda = 0.746 \pm 0.009$~\cite{ParticleDataGroup:2024cfk}.}
	\label{fig:compare_compass}
\end{figure}

On the other hand, the TF effect on the spin transfer suppression should reduce if the measurement is carried out at a higher energy. In Fig.~\ref{fig:compare_compass}, the results are compared with the COMPASS data~\cite{COMPASS:2009nhs}, in which the $D_{LL}$ for both $\Lambda$ and $\bar\Lambda$ are reported. As expected, the suppression by the TF, shown by the difference between the dashed curve and the solid curve, is much smaller. The results including the TF contribution are still much greater than the data of $\Lambda$. We should note that only the leading channel is taken into account in the estimation of TF, which may be understood as a lower limit of the TF contribution. Since the energy of the COMPASS experiment is higher than those of HERMES and CLAS, contributions beyond the leading channel may also have non-negligible effects. 
	
In addition, one can observe in Fig.~\ref{fig:compare_compass} that the $D_{LL}$ data of $\bar\Lambda$ can be well described with only CF contributions. This confirms the physical picture as presented above from a different aspect. Since the $\bar\Lambda$ contains three valence antiquarks, the CF is inhibited by requiring the generation of additional quark-antiquark pair if from a struck quark or by the sea quark density in the nucleon if from a struck antiquark. For the TF, the probability is further reduced by antiquark densities in the nucleon, since an antiquark pair is required if to compete with the CF during the hadronization process. So the TF effects on $\bar\Lambda$ production from nucleon targets are negligible.

	
{\it Summary and outlook}---For $\Lambda$ production in existing fixed-target SIDIS experiments, there is not a clear isolation between CF and TF contributions. It is important to take into account both mechanisms, especially for the understanding of spin-related observables. 
	
Taking the longitudinal spin transfer $D_{LL}$, we demonstrate that the experimentally measured value can be significantly suppressed by the TF contribution. With the FrF estimated by a simple model, we show that the inclusion of TF can much alleviate the tension between experimental data and CF predictions. Even if at very large $x_F$ or $z$, the TF effect on $D_{LL}$ can still be huge. Therefore, $D_{LL}$ is a sensitive observable to the origin of the produced $\Lambda$, shedding light on the hadronization dynamics.  

The TF contributions at current experiments at JLab are very significant. It provides opportunities to systematically investigate the TF effects, taking advantage of spin-related observables.
The presented spin-transfer effect on $\Lambda$ production can be extended to other hyperons, which may also have feed-down contributions to the $\Lambda$ measurements and deserve systematic studies in future.
Global analysis of FFs, especially spin-dependent ones, including SIDIS data should carefully handle TF contributions.
Measuring azimuthal asymmetries that are absent in the CF region also offer a potential method to probe the TF contribution, and such measurements could be pursued in future experiments.
The TF effects generally decrease with increasing energies of the experiment, which can be tested at future EICs~\cite{Accardi:2012qut,AbdulKhalek:2021gbh,Anderle:2021wcy}.
It may also shed light on the understanding of the spin-transfer measurements in polarized $pp$ collisions~\cite{STAR:2009hex, STAR:2018pps, STAR:2023hwu}. More scrutiny, combining various high-energy reaction processes, is desired on target fragmentations.

\acknowledgments{
We thank Chao-Hsi Chang, Yongjie Deng, Zhe Zhang, and Jing Zhao for the valuable discussions. This work was supported by the National Key R\&D Program of China No.~2024YFA1611004 and by the National Natural Science Foundation of China\ (Grant No. 12175117, No. 12321005 and No. 12475084) and Shandong Province Natural Science Foundation\ (Grant No. ZFJH202303 and No. ZR2024MA012).
}


\begin{thebibliography}{}
		
\bibitem{Bunce:1976yb}
G.~Bunce \textit{et al.},
$\Lambda^0$ hyperon polarization in inclusive production by 300-GeV protons on beryllium,
\href{https://doi.org/10.1103/PhysRevLett.36.1113}{Phys. Rev. Lett. \textbf{36}, 1113 (1976)}.
		
\bibitem{Kane:1978nd}
G.~L.~Kane, J.~Pumplin, and W.~Repko,
Transverse quark polarization in large-$p_T$ reactions, $e^+ e^-$ jets, and leptoproduction: A test of quantum chromodynamics,
\href{https://doi.org/10.1103/PhysRevLett.41.1689}{Phys. Rev. Lett. \textbf{41}, 1689 (1978)}.
		
\bibitem{Lee:1957qs}
T.~D.~Lee and C.~N.~Yang,
General partial wave analysis of the decay of a hyperon of spin 1/2,
\href{https://doi.org/10.1103/PhysRev.108.1645}{Phys. Rev. \textbf{108}, 1645 (1957)}.
		
\bibitem{Metz:2016swz}
A.~Metz and A.~Vossen,
Parton fragmentation functions,
\href{https://doi.org/10.1016/j.ppnp.2016.08.003}{Prog. Part. Nucl. Phys. \textbf{91}, 136 (2016)}.
		
\bibitem{Chen:2023kqw}
K.~B.~Chen, T.~Liu, Y.~K.~Song and S.~Y.~Wei,
Several topics on transverse momentum-dependent fragmentation functions,
\href{https://doi.org/10.3390/particles6020029}{Particles \textbf{6}, 515 (2023)}.
		
\bibitem{ALEPH:1996oew}
D.~Buskulic \textit{et al.} (ALEPH Collaboration),
Measurement of $\Lambda$ polarization from Z decays,
\href{https://doi.org/10.1016/0370-2693(96)00300-0}{Phys. Lett. B \textbf{374}, 319 (1996)}.
		
\bibitem{OPAL:1997oem}
K.~Ackerstaff \textit{et al.} (OPAL Collaboration),
Polarization and forward-backward asymmetry of $\Lambda$ baryons in hadronic $Z^0$ decays,
\href{https://doi.org/10.1007/s100520050123}{Eur. Phys. J. C \textbf{2}, 49 (1998)}.
		
\bibitem{Belle:2018ttu}
Y.~Guan \textit{et al.} (Belle Collaboration),
Observation of transverse $\Lambda/\bar{\Lambda}$ hyperon polarization in $e^+e^-$ annihilation at Belle,
\href{https://doi.org/10.1103/PhysRevLett.122.042001}{Phys. Rev. Lett. \textbf{122}, 042001 (2019)}.
		
\bibitem{E665:1999fso}
M.~R.~Adams \textit{et al.} (E665 Collaboration),
$\Lambda$ and $\bar\Lambda$ polarization from deep inelastic muon scattering,
\href{https://doi.org/10.1007/s100520000493}{Eur. Phys. J. C \textbf{17}, 263 (2000)}.
		
\bibitem{HERMES:1999buc}
A.~Airapetian \textit{et al.} (HERMES Collaboration),
Measurement of longitudinal spin transfer to $\Lambda$ hyperons in deep inelastic lepton scattering,
\href{https://doi.org/10.1103/PhysRevD.64.112005}{Phys. Rev. D \textbf{64}, 112005 (2001)}.
		
\bibitem{NOMAD:2000wdf}
P.~Astier \textit{et al.} (NOMAD Collaboration),
Measurement of the $\Lambda$ polarization in $\nu_\mu$ charged current interactions in the NOMAD experiment,
\href{https://doi.org/10.1016/S0550-3213(00)00503-4}{Nucl. Phys. \textbf{B588}, 3 (2000)}.

\bibitem{ZEUS:2006ugq}
S.~Chekanov \textit{et al.} (ZEUS Collaboration),
Measurement of $K^0_{S}$, $\Lambda$ and $\bar{\Lambda}$ production at HERA,
\href{https://doi.org/10.1140/epjc/s10052-007-0299-2}{Eur. Phys. J. C \textbf{51}, 1 (2007)}.
		
\bibitem{HERMES:2006lro}
A.~Airapetian \textit{et al.} (HERMES Collaboration),
Longitudinal spin transfer to the $\Lambda$ Hyperon in semiinclusive deep-inelastic scattering,
\href{https://doi.org/10.1103/PhysRevD.74.072004}{Phys. Rev. D \textbf{74}, 072004 (2006)}.
		
\bibitem{COMPASS:2009nhs}
M.~Alekseev \textit{et al.} (COMPASS Collaboration),
Measurement of the longitudinal spin transfer to $\Lambda$ and $\bar\Lambda$ hyperons in polarised muon DIS,
\href{https://doi.org/10.1140/epjc/s10052-009-1143-7}{Eur. Phys. J. C \textbf{64}, 171 (2009)}.
		
\bibitem{COMPASS:2021bws}
G.~D.~Alexeev \textit{et al.} (COMPASS Collaboration),
Probing transversity by measuring \ensuremath{\Lambda} polarisation in SIDIS,
\href{https://doi.org/10.1016/j.physletb.2021.136834}{Phys. Lett. B \textbf{824}, 136834 (2022)}.
		
\bibitem{ATLAS:2014ona}
G.~Aad \textit{et al.} (ATLAS Collaboration),
Measurement of the transverse polarization of $\Lambda$ and $\bar{\Lambda}$ hyperons produced in proton-proton collisions at $\sqrt{s}=7$ TeV using the ATLAS detector,
\href{https://doi.org/10.1103/PhysRevD.91.032004}{Phys. Rev. D \textbf{91}, 032004 (2015)}.
		
\bibitem{STAR:2018fqv}
J.~Adam \textit{et al.} (STAR Collaboration),
Transverse spin transfer to $\Lambda$ and $\bar{\Lambda}$ hyperons in polarized proton-proton collisions at $\sqrt{s}=200\,\mathrm{GeV}$,
\href{https://doi.org/10.1103/PhysRevD.98.091103}{Phys. Rev. D \textbf{98}, 091103 (2018)}.
		
\bibitem{STAR:2009hex}
B.~I.~Abelev \textit{et al.} (STAR Collaboration),
Longitudinal spin transfer to $\Lambda$ and $\bar\Lambda$ hyperons in polarized proton-proton collisions at $\sqrt{s} = 200~\rm GeV$,
\href{https://doi.org/10.1103/PhysRevD.80.111102}{Phys. Rev. D \textbf{80}, 111102 (2009)}.
		
\bibitem{STAR:2018pps}
J.~Adam \textit{et al.} (STAR Collaboration),
Improved measurement of the longitudinal spin transfer to $\Lambda$ and $\bar \Lambda$ hyperons in polarized proton-proton collisions at $\sqrt s$ = 200 GeV,
\href{https://doi.org/10.1103/PhysRevD.98.112009}{Phys. Rev. D \textbf{98}, 112009 (2018)}.
		
\bibitem{STAR:2023hwu}
M.~Abdulhamid \textit{et al.} (STAR Collaboration),
Longitudinal and transverse spin transfer to $\Lambda$ and $\bar\Lambda$ hyperons in polarized p+p collisions at $\sqrt s$=200\,GeV,
\href{https://doi.org/10.1103/PhysRevD.109.012004}{Phys. Rev. D \textbf{109}, 012004 (2024)}.
		
\bibitem{Gell-Mann:1964ewy}
M.~Gell-Mann,
A schematic model of baryons and mesons,
\href{https://doi.org/10.1016/S0031-9163(64)92001-3}{Phys. Lett. \textbf{8}, 214 (1964)}.
		
\bibitem{Zweig:1964ruk}
G. Zweig, An SU(3) model for strong interaction symmetry and its breaking, Version 1, Technical Report CERN-TH-401, CERN, Geneva, 1964.
		
\bibitem{Ellis:1995fc}
J.~R.~Ellis, D.~Kharzeev, and A.~Kotzinian,
The proton spin puzzle and $\Lambda$ polarization in deep-inelastic scattering,
\href{https://doi.org/10.1007/BF02907428}{Z. Phys. C \textbf{69}, 467 (1996)}.
		
\bibitem{Lu:1995np}
W.~Lu and B.-Q.~Ma,
The strange quark spin of the proton in semi-inclusive $\Lambda$ leptoproduction,
\href{https://doi.org/10.1016/0370-2693(95)00927-D}{Phys. Lett. B \textbf{357}, 419 (1995)}.

\bibitem{Brodsky:1996hc}
S.~J.~Brodsky and B.-Q.~Ma,
The quark-antiquark asymmetry of the nucleon sea,
\href{https://doi.org/10.1016/0370-2693(96)00597-7}{Phys. Lett. B \textbf{381}, 317 (1996)}.

\bibitem{Jaffe:1996wp}
R.~L.~Jaffe,
Polarized $\Lambda$'s in the current fragmentation region,
\href{https://doi.org/10.1103/PhysRevD.54.R6581}{Phys. Rev. D \textbf{54}, R6581 (1996)}.
		
\bibitem{Ma:2000uv}
B.-Q.~Ma, I.~Schmidt, J.~Soffer, and J.~J.~Yang,
The quark-antiquark asymmetry of the nucleon sea from $\Lambda$ and $\bar\Lambda$ fragmentation,
\href{https://doi.org/10.1016/S0370-2693(00)00906-0}{Phys. Lett. B \textbf{488}, 254 (2000)}.
		
\bibitem{Ellis:2007ig}
J.~R.~Ellis, A.~Kotzinian, D.~Naumov, and M.~Sapozhnikov,
Longitudinal polarisation of $\Lambda$ and $\bar\Lambda$ hyperons in lepton-nucleon deep-inelastic scattering,
\href{https://doi.org/10.1140/epjc/s10052-007-0381-9}{Eur. Phys. J. C \textbf{52}, 283 (2007)}.
		
\bibitem{Chi:2014xba}
Y.~Chi, X.~Du and B.-Q.~Ma,
Nucleon strange $s\bar{s}$ asymmetry to the $\Lambda/\bar{\Lambda}$ fragmentation,
\href{https://doi.org/10.1103/PhysRevD.90.074003}{Phys. Rev. D \textbf{90}, 074003 (2014)}.
		
\bibitem{Du:2017nzy}
X.~Du and B.-Q.~Ma,
Strange quark-antiquark asymmetry of nucleon sea from $\Lambda/\bar\Lambda$ polarization,
\href{https://doi.org/10.1103/PhysRevD.95.014029}{Phys. Rev. D \textbf{95}, 014029 (2017)}.
		
\bibitem{Berman:1971xz}
S.~M.~Berman, J.~D.~Bjorken, and J.~B.~Kogut,
Inclusive processes at high transverse momentum,
\href{https://doi.org/10.1103/PhysRevD.4.3388}{Phys. Rev. D \textbf{4}, 3388 (1971)}.
		
\bibitem{Collins:1981uw}
J.~C.~Collins and D.~E.~Soper,
Parton distribution and decay functions,
\href{https://doi.org/10.1016/0550-3213(82)90021-9}{Nucl. Phys. \textbf{B194}, 445 (1982)}.
		
\bibitem{Trentadue:1993ka}
L.~Trentadue and G.~Veneziano,
Fracture functions: An improved description of inclusive hard processes in QCD,
\href{https://doi.org/10.1016/0370-2693(94)90292-5}{Phys. Lett. B \textbf{323}, 201 (1994)}.
		
\bibitem{Berera:1995fj}
A.~Berera and D.~E.~Soper,
Behavior of diffractive parton distribution functions,
\href{https://doi.org/10.1103/PhysRevD.53.6162}{Phys. Rev. D \textbf{53}, 6162 (1996)}.
		
\bibitem{Grazzini:1997rw}
M.~Grazzini,
Semi-inclusive DIS: An explicit calculation in the target fragmentation region,
\href{https://doi.org/10.1016/S0920-5632(97)01052-9}{Nucl. Phys. B, Proc. Suppl. \textbf{64}, 147 (1998)}.
		
\bibitem{deFlorian:1997wi}
D.~de Florian and R.~Sassot,
Phenomenology of forward hadrons in deep inelastic scattering: Fracture functions and its ${Q}^{2}$ evolution,
\href{https://doi.org/10.1103/PhysRevD.56.426}{Phys. Rev. D \textbf{56}, 426 (1997)}.
		
\bibitem{Grazzini:1999vz}
M.~Grazzini, G.~M.~Shore, and B.~E.~White,
Target fragmentation in semi-inclusive DIS: Fracture functions, cut vertices and the OPE,
\href{https://doi.org/10.1016/S0550-3213(99)00347-8}{Nucl. Phys. \textbf{B555}, 259 (1999)}.
		
\bibitem{Meng:1991da}
R.~b.~Meng, F.~I.~Olness, and D.~E.~Soper,
Semi-inclusive deeply inelastic scattering at electron-proton colliders,
\href{https://doi.org/10.1016/0550-3213(92)90230-9}{Nucl. Phys. \textbf{B371}, 79 (1992)}.
		
\bibitem{Meng:1995yn}
R.~Meng, F.~I.~Olness, and D.~E.~Soper,
Semi-inclusive deeply inelastic scattering at small $q_T$,
\href{https://doi.org/10.1103/PhysRevD.54.1919}{Phys. Rev. D \textbf{54}, 1919 (1996)}.
		
\bibitem{Nadolsky:1999kb}
P.~M.~Nadolsky, D.~R.~Stump, and C.-P.~Yuan,
Semi-inclusive hadron production at DESY HERA: The effect of QCD gluon resummation,
\href{https://doi.org/10.1103/PhysRevD.61.014003}{Phys. Rev. D \textbf{61}, 014003 (1999)}
[\href{https://doi.org/10.1103/PhysRevD.64.059903}{Phys. Rev. D \textbf{64}, 059903(E) (2001)}].
		
\bibitem{Ji:2004xq}
X.~Ji, J.-P.~Ma, and F.~Yuan,
QCD factorization for spin-dependent cross sections in DIS and Drell-Yan processes at low transverse momentum,
\href{https://doi.org/10.1016/j.physletb.2004.07.026}{Phys. Lett. B \textbf{597}, 299 (2004)}.
		
\bibitem{Ji:2004wu}
X.~Ji, J.-P.~Ma, and F.~Yuan,
QCD factorization for semi-inclusive deep-inelastic scattering at low transverse momentum,
\href{https://doi.org/10.1103/PhysRevD.71.034005}{Phys. Rev. D \textbf{71}, 034005 (2005)}.
		
\bibitem{Koike:2006fn}
Y.~Koike, J.~Nagashima, and W.~Vogelsang,
Resummation for polarized semi-inclusive deep-inelastic scattering at small transverse momentum,
\href{https://doi.org/10.1016/j.nuclphysb.2006.03.009}{Nucl. Phys. \textbf{B744}, 59 (2006)}.
		
\bibitem{Collins:2011zzd}
J.~Collins,
{\it Foundations of Perturbative QCD},
Cambridge Monographs on Particle Physics, Nuclear Physics and Cosmology Series Vol. 32 (Cambridge University Press, Cambridge, England, 2011).
		
\bibitem{Collins:2017oxh}
J.~Collins and T.~C.~Rogers,
Connecting different TMD factorization formalisms in QCD,
\href{https://doi.org/10.1103/PhysRevD.96.054011}{Phys. Rev. D \textbf{96}, 054011 (2017)}.
		
\bibitem{Ceccopieri:2007th}
F.~A.~Ceccopieri and L.~Trentadue,
A new fracture function approach to QCD initial state radiation,
\href{https://doi.org/10.1016/j.physletb.2007.07.074}{Phys. Lett. B \textbf{655}, 15 (2007)}.
		
\bibitem{Anselmino:2011ss}
M.~Anselmino, V.~Barone, and A.~Kotzinian,
SIDIS in the target fragmentation region: Polarized and transverse momentum dependent fracture functions,
\href{https://doi.org/10.1016/j.physletb.2011.03.067}{Phys. Lett. B \textbf{699}, 108 (2011)}.
		
\bibitem{Chai:2019ykk}
X.~P.~Chai, K.~B.~Chen, J.~P.~Ma, and X.~B.~Tong,
Fracture functions in different kinematic regions and their factorizations,
\href{https://doi.org/10.1007/JHEP10(2019)285}{J. High Energy Phys. 10 (2019) 285}.
		
\bibitem{Yang:2020sos}
W.~Yang and F.~Huang,
Deep inelastic scattering in the target fragmentation region,
\href{https://doi.org/10.1142/S0217751X20502127}{Int. J. Mod. Phys. A \textbf{35}, 2050212 (2020)}.
		
\bibitem{Chen:2021vby}
K.~B.~Chen, J.~P.~Ma, and X.~B.~Tong,
Matching of fracture functions for SIDIS in target fragmentation region,
\href{https://doi.org/10.1007/JHEP11(2021)038}{J. High Energy Phys. 11 (2021) 038}.
		
\bibitem{Chen:2023wsi}
K.~B.~Chen, J.~P.~Ma, and X.~B.~Tong,
Twist-3 contributions in semi-inclusive DIS in the target fragmentation region,
\href{https://doi.org/10.1103/PhysRevD.108.094015}{Phys. Rev. D \textbf{108}, 094015 (2023)}.

\bibitem{Chen:2024brp}
K.~B.~Chen, J.~P.~Ma, and X.~B.~Tong,
Gluonic contributions to semi-inclusive DIS in the target fragmentation region,
\href{https://doi.org/10.1007/JHEP05(2024)298}{J. High Energy Phys. 05 (2024) 298}.
		
\bibitem{ZEUS:1993vio}
M.~Derrick \textit{et al.} (ZEUS Collaboration),
Observation of events with a large rapidity gap in deep inelastic scattering at HERA,
\href{https://doi.org/10.1016/0370-2693(93)91645-4}{Phys. Lett. B \textbf{315}, 481 (1993)}.

\bibitem{Ammosov:1980zs}
V.~V.~Ammosov \textit{et al.},
Charged current events with neutral strange particles in high-energy antineutrino interactions,
\href{https://doi.org/10.1016/0550-3213(81)90173-5}{Nucl. Phys. \textbf{B177}, 365 (1981)}.

\bibitem{Birmingham-Bonn-CERN-London-Munich-Oxford:1985ret}
G.~T.~Jones \textit{et al.} (Birmingham-Bonn-CERN-London-Munich-Oxford Collaboration),
Polarization of $\Lambda$ hyperons produced inclusively in $\nu p$ and $\bar\nu p$ charged current interactions,
\href{https://doi.org/10.1007/BF01550245}{Z. Phys. C \textbf{28}, 23 (1985)}.

\bibitem{WA59:1991cop}
S.~Willocq \textit{et al.} (WA59 Collaboration),
Neutral strange particle production in antineutrino-neon charged current interactions,
\href{https://doi.org/10.1007/BF01597556}{Z. Phys. C \textbf{53}, 207 (1992)}.

\bibitem{Jones:1992bm}
G.~T.~Jones \textit{et al.} (WA21 Collaboration),
Neutral strange particle production in neutrino and antineutrino charged current interactions on protons,
\href{https://doi.org/10.1007/BF01565049}{Z. Phys. C \textbf{57}, 197 (1993)}.

\bibitem{NOMAD:2001iup}
P.~Astier \textit{et al.} (NOMAD Collaboration),
Measurement of the $\bar\Lambda$ polarization in $\nu_\mu$ charged current interactions in the NOMAD experiment,
\href{https://doi.org/10.1016/S0550-3213(01)00181-X}{Nucl. Phys. \textbf{B605}, 3 (2001)}.

\bibitem{E632:1994njp}
D.~DeProspo \textit{et al.} (E632 Collaboration),
Neutral strange particle production in neutrino and antineutrino charged-current interactions on neon,
\href{https://doi.org/10.1103/PhysRevD.50.6691}{Phys. Rev. D \textbf{50}, 6691 (1994)}.

\bibitem{EuropeanMuon:1984ucw}
M.~Arneodo \textit{et al.} (European Muon Collaboration),
Quark and diquark fragmentation into neutral strange particles as observed in muon-proton interactions at 280 GeV,
\href{https://doi.org/10.1016/0370-2693(84)90969-9}{Phys. Lett. \textbf{145}B, 156 (1984)}.
        
\bibitem{Berger:1987zu}
E.~L.~Berger,
Semiinclusive inelastic electron scattering from nuclei,
\textit{Proceedings of the Workshop on Electronuclear Physics with Internal Targets} (1987), ANL-HEP-CP-87-45.
		
\bibitem{Mulders:2000jt}
P.~J.~Mulders,
Current fragmentation in semiinclusive leptoproduction,
\href{https://doi.org/10.1063/1.1413147}{AIP Conf. Proc. \textbf{588}, 75 (2001)}.
		
\bibitem{Boglione:2016bph}
M.~Boglione, J.~Collins, L.~Gamberg, J.~O.~Gonzalez-Hernandez, T.~C.~Rogers, and N.~Sato,
Kinematics of current region fragmentation in semi-inclusive deeply inelastic scattering,
\href{https://doi.org/10.1016/j.physletb.2017.01.021}{Phys. Lett. B \textbf{766}, 245 (2017)}.
		
\bibitem{Boglione:2019nwk}
M.~Boglione, A.~Dotson, L.~Gamberg, S.~Gordon, J.~O.~Gonzalez-Hernandez, A.~Prokudin, T.~C.~Rogers, and N.~Sato,
Mapping the kinematical regimes of semi-inclusive deep inelastic scattering,
\href{https://doi.org/10.1007/JHEP10(2019)122}{J. High Energy Phys. 10 (2019) 122}.
		
\bibitem{Belostotski:2011zza}
S.~Belostotski, D.~Veretennikov, and Yu.~Naryshkin,
Spin transfer coefficient $D^\Lambda_{LL}$ to $\Lambda$ hyperon in semi-inclusive DIS at HERMES,
\href{https://doi.org/10.1088/1742-6596/295/1/012114}{J. Phys. Conf. Ser. \textbf{295}, 012114 (2011)}.
		
\bibitem{McEneaney:2022bsf}
M.~McEneaney (CLAS Collaboration),
Longitudinal spin transfer to $\Lambda$ hyperons in CLAS12,
\href{https://doi.org/10.7566/JPSCP.37.020304}{JPS Conf. Proc. \textbf{37}, 020304 (2022)}.

\bibitem{McEneaney:2023s}
M. McEneaney (CLAS Collaboration),
Longitudinal spin transfer to $\Lambda$ hyperons with CLAS12, reported at
\textit{The 25th International Spin Symposium (SPIN 2023), Durham, NC, USA} (2023)
\href{https://indico.jlab.org/event/663/contributions/13283/}{https://indico.jlab.org/event/663/contributions/13283/}
		
\bibitem{Dudek:2012vr}
J.~Dudek \textit{et al.},
Physics opportunities with the 12 GeV upgrade at Jefferson Lab,
\href{https://doi.org/10.1140/epja/i2012-12187-1}{Eur. Phys. J. A \textbf{48}, 187 (2012)}.
		
\bibitem{Accardi:2012qut}
A.~Accardi, J.L. Albacete, M. Anselmino \textit{et al.},
Electron-Ion Collider: The next QCD frontier,
\href{https://doi.org/10.1140/epja/i2016-16268-9}{Eur. Phys. J. A \textbf{52}, 268 (2016)}.
		
\bibitem{AbdulKhalek:2021gbh}
R.~Abdul Khalek \textit{et al.},
Science requirements and detector concepts for the electron-ion collider: EIC Yellow Report,
\href{https://doi.org/10.1016/j.nuclphysa.2022.122447}{Nucl. Phys. \textbf{A1026}, 122447 (2022)}.
		
\bibitem{Anderle:2021wcy}
D.~P.~Anderle \textit{et al.},
Electron-ion collider in China,
\href{https://doi.org/10.1007/s11467-021-1062-0}{Front. Phys. (Beijing) \textbf{16}, 64701 (2021)}.

\bibitem{Boer:1999uu}
D.~Boer, R.~Jakob, and P.~J.~Mulders,
Angular dependences in electroweak semi-inclusive leptoproduction,
\href{https://doi.org/10.1016/S0550-3213(99)00586-6}{Nucl. Phys. \textbf{B564}, 471 (2000)}.
		
\bibitem{Bacchetta:2006tn}
A.~Bacchetta, M.~Diehl, K.~Goeke, A.~Metz, P.~J.~Mulders and M.~Schlegel,
Semi-inclusive deep inelastic scattering at small transverse momentum,
\href{https://iopscience.iop.org/article/10.1088/1126-6708/2007/02/093}{J. High Energy Phys. 02 (2007) 093.}
		
\bibitem{Yang:2016qsf}
Y.~Yang and Z.~Lu,
Polarized \ensuremath{\Lambda} hyperon production in semi-inclusive deep inelastic scattering off an unpolarized nucleon target,
\href{https://doi.org/10.1103/PhysRevD.95.074026}{Phys. Rev. D \textbf{95}, 074026 (2017)}.

\bibitem{Zhao:2024zpy}
J.~Zhao, Z.~Zhang, Z.~t.~Liang, T.~Liu and Y.~j.~Zhou,
Semi-inclusive production of spin-3/2 hadrons in deep inelastic scattering,
\href{https://doi.org/10.1103/PhysRevD.109.074017}{Phys. Rev. D \textbf{109}, 074017 (2024)}.
		
\bibitem{Farrar:1975yb}
G.~R.~Farrar and D.~R.~Jackson,
Pion and nucleon structure functions near $x\,$=1,
\href{https://doi.org/10.1103/PhysRevLett.35.1416}{Phys. Rev. Lett. \textbf{35}, 1416 (1975)}.
		
\bibitem{Brodsky:1994kg}
S.~J.~Brodsky, M.~Burkardt, and I.~Schmidt,
QCD constraints on the shape of polarized quark and gluon distributions,
\href{https://doi.org/10.1016/0550-3213(95)00009-H}{Nucl. Phys. \textbf{B441}, 197 (1995)}.
		
\bibitem{deFlorian:1997zj}
D.~de Florian, M.~Stratmann, and W.~Vogelsang,
QCD analysis of unpolarized and polarized $\Lambda$-baryon production in leading and next-to-leading order,
\href{https://doi.org/10.1103/PhysRevD.57.5811}{Phys. Rev. D \textbf{57}, 5811 (1998)}.
		
\bibitem{Bacchetta:2008af}
A.~Bacchetta, F.~Conti, and M.~Radici,
Transverse-momentum distributions in a diquark spectator model,
\href{https://doi.org/10.1103/PhysRevD.78.074010}{Phys. Rev. D \textbf{78}, 074010 (2008)}.
		
\bibitem{Jimenez-Delgado:2014twa}
P.~Jimenez-Delgado and E.~Reya,
Delineating parton distributions and the strong coupling,
\href{https://doi.org/10.1103/PhysRevD.89.074049}{Phys. Rev. D \textbf{89}, 074049 (2014)}.

\bibitem{Anselmino:2013lza}
M.~Anselmino, M.~Boglione, J.~O.~Gonzalez Hernandez, S.~Melis and A.~Prokudin,
Unpolarised transverse momentum dependent distribution and fragmentation functions from SIDIS multiplicities,
\href{https://doi.org/10.1007/JHEP04(2014)005}{J. High Energy Phys. 04 (2014) 005}.
		
\bibitem{Callos:2020qtu}
D.~Callos, Z.~B.~Kang and J.~Terry,
Extracting the transverse momentum dependent polarizing fragmentation functions,
\href{https://doi.org/10.1103/PhysRevD.102.096007}{Phys. Rev. D \textbf{102}, 096007 (2020)}.

\bibitem{ParticleDataGroup:2024cfk}
S.~Navas \textit{et al.} (Particle Data Group),
Review of particle physics,
\href{https://doi.org/10.1103/PhysRevD.110.030001}{Phys. Rev. D \textbf{110}, 030001 (2024)}.

\bibitem{SupplementalMaterial}
See Supplemental Material at 
\href{http://link.aps.org/supplemental/10.1103/1nd6-9nw9}{http://link.aps.org/supplemental/10.1103/1nd6-9nw9}
differential cross section of target fragmentation, its distribution and the spin transfer at large $x_F$.

		
		
\end{thebibliography}
\end{document}